\documentclass[10pt,aps,prl,twocolumn,notitlepage,showpacs,superscriptaddress,groupedaddress]{revtex4-1} 
\usepackage{amsmath}
\usepackage{amssymb}
\usepackage[mathscr]{eucal}
\usepackage[T1]{fontenc}
\usepackage{graphicx}
\usepackage{lipsum}

\usepackage{color}
\definecolor{dgreen}{rgb}{0,0.7,0}
\def\redw#1{{\color{black} #1}}

\usepackage{url,hyperref}

\newcommand{\ee}{\mathrm{e}}

\newcommand{\iu}{{\mathrm{i}\mkern1mu}}

\newcommand{\eqa}[1]{\begin{align}#1\end{align}}

	\newcommand{\ket}[1]{ |\;#1\;\rangle}

	\newcommand{\iprod}[2]{ \langle\;#1\;|\;#2\;\rangle}

	\newcommand{\mb}[1]{\mathbb{#1}}
	
	\newcommand{\dx}[1]{\hat{#1}}
	\newcommand{\nn}{\nonumber}
	\newcommand{\la}{\langle}
	\newcommand{\ra}{\rangle}
	\newcommand{\p}{\partial}
	
	\newcommand{\mc}[1]{\mathcal{#1}}

\def\be{\begin{equation}}
\def\ee{\end{equation}}
\def\bea{\begin{eqnarray}}
\def\eea{\end{eqnarray}}
\def\bed{\begin{dmath*}}
\def\eed{\end{dmath*}}

\begin{document}
	\title{Anomalous heat equation in a system connected to thermal reservoirs}
	\author{Priyanka}\email{ priyanka@icts.res.in}
	\affiliation{International Centre for Theoretical Sciences, Tata Institute of Fundamental Research, Bengaluru 560089, India}
	\author{Aritra Kundu}\email{ aritrak@icts.res.in}
	\affiliation{International Centre for Theoretical Sciences, Tata Institute of Fundamental Research, Bengaluru 560089, India}
	\author{Abhishek Dhar}\email{abhishek.dhar@icts.res.in}
	\affiliation{International Centre for Theoretical Sciences, Tata Institute of Fundamental Research, Bengaluru 560089, India}
	\author{Anupam Kundu}\email{anupam.kundu@icts.res.in  }
	\affiliation{International Centre for Theoretical Sciences, Tata Institute of Fundamental Research, Bengaluru 560089, India}
	
	\date{\today}

	\begin{abstract}
		\noindent 
		{
			We study anomalous transport in a one-dimensional system with two conserved quantities in presence of thermal baths. In this system we derive exact expressions of the temperature profile and the two point correlations in steady state as well as in the non-stationary state where the later describe the relaxation to the steady state. In contrast to the Fourier heat equation in the diffusive case, here we show that the evolution of the temperature profile is governed by a non-local anomalous heat equation. We provide numerical verifications  of our results.
		}
	\end{abstract}
	
	\maketitle
	
	\noindent
	Transport of energy across an extended system is a paradigm of the vast class of non-equilibrium phenomena. 
	At a macroscopic level this phenomena is often described by the phenomenological Fourier's law which 
	relates the energy current density $j(x,t)$ to the gradient of the temperature field $T(x,t)$: $j = -\kappa~\partial_x T$ where $\kappa$ is the thermal conductivity. This law 
	implies diffusive energy flow across the system described by the Fourier heat equation 
	\begin{equation}
	\partial_\tau T(x,t)= D~\partial_x^2 T(x,t), \label{heat-diff}
	\end{equation}
	where $D=\kappa/c$ (for simplicity we assume $\kappa$ and the specific heat $c$ to be independent of temperature). This equation is widely used in experiments to understand the spreading of local energy perturbations in equilibrium as well as the non-equilibrium dynamics of systems connected to reservoirs.

	Surprisingly, several theoretical \cite{Narayan2002, VanBeijeren2012, Spohn2014}, numerical as well as experimental studies \cite{Lee2017} suggest that in many one and two dimensional systems heat transfer is anomalous in the sense that Fourier's law is not valid \cite{Lepri2003a, Dhar2008, Lepri2016}. This phenomenon is usually manifested by several interesting features like divergence of thermal conductivity $\kappa$ with system size  $L$ as $\kappa \sim L^\alpha;~0< \alpha <1$, power-law decay of the equilibrium current-current auto-correlations, super-diffusive spreading of local energy perturbations, nonlinear stationary temperature profiles (even for small temperature differences) and the presence of boundary singularities in these profiles~\cite{Dhar2001,Dhar2008, Lepri2016, Narayan2002, Lepri1998, Liu2014, Das2014, Bernardin2015, Bernardin2012, Jara2015}.

	
	There is currently no  general framework to describe and explain anomalous heat transport. 
	Recently, the theory of nonlinear fluctuating hydrodynamics has been remarkably successful in predicting anomalous scaling of dynamical correlations of conserved quantities in one-dimensional Hamiltonian systems and the corresponding slow decay of the equilibrium current-current autocorrelations  \cite{VanBeijeren2012, Spohn2014, Spohn2016, Mendl2015}. This approach provides diverging thermal conductivity (via Green-Kubo formula) as well as L\'evy scaling for the spreading of local energy perturbation. On the rigorous side, computations were done for a model of harmonic chain whose Hamiltonian dynamics was supplemented by a stochastic part that kept the conservation laws (number, energy, momentum) intact --- we refer to this model as the harmonic chain momentum exchange (HCME) model. For the infinite HCME system it was shown exactly that the current autocorrelation has a $\sim t^{-1/2}$ decay \cite{Basile2006a}. 
	It was also shown that, in contrast to Eq.~\eqref{heat-diff}, the evolution of an initially localized energy perturbation satisfied a non-local fractional diffusion equation  
	$\partial_t e(x,t) = -c(-\Delta)^{3/4} e(x,t)$,
	where $e(x,t)$ is the energy perturbation and  $c$ is some constant \cite{Bernardin2015, Bernardin2012, Jara2015, Bernardin2016}. The fractional laplacian operator 
	$(-\Delta)^{3/4}$ in the infinite space is defined by its Fourier spectrum: $|q|^{3/2}$ whereas the same for the normal Laplacian operator $-\Delta \equiv -\partial^2_x$ is $q^2$. In  real space $(-\Delta)^{3/4}$ operator is non-local~\cite{fractional1, fractional2}. 
	
	While all these studies consider transport in {\it isolated} systems, quite often the transport set up in an experiment consists of an extended system connected at the two ends to heat baths at different temperatures.  For diffusive systems Eq.~\eqref{heat-diff} continues to describe both non-equilibrium steady state (NESS) and time-dependent properties in this set-up. It is then natural  to ask: what would be the corresponding evolution equation for the temperature profile in the case of anomalous transport in the experimental set-up? A major problem that now arises  follows from the fact that the 
	fractional Laplacian is a non-local operator and so extending its definition 
	to a finite domain is non-trivial. Several studies have addressed this issue, using a phenomenological approach,  in the context of Levy walks and Levy flights in finite domains \cite{Viswanathan2000,Zoia2007}. 
	It is thus crucial to have examples of   specific microscopic models  of systems exhibiting anomalous transport, for which the time-evolution  equation in an open system set up can be  derived analytically, and where one  can  
	see the non-local and fractional equation forms explicitly. This is the main aim of this Letter. 
	
	Such attempts have recently been made in \cite{Lepri2009, Delfini2008, Cividini2017} where the problem of non-linear steady state temperature profiles and their time-evolution in  the HCME model was addressed.  The specific model studied was a harmonic chain of $N$ particles where,  in  addition to the  Hamiltonian dynamics, the momenta of nearest neighborhood particles is exchanged randomly at a constant rate $\gamma$. 
	The chain is attached to two Langevin baths at the two ends at temperatures $T_\ell$ and $T_r$. This system has three conserved fields: the stretch $r_i=q_{i+1}-q_i$  (where $q_i,~i=1,\ldots,N$ are the particle positions) the momentum, $p_i$ and the energy $\epsilon_i=p_i^2/2+r_i^2/2$. 
	This system shows anomalous current behavior $j \sim N^{-1/2}$ as well as exhibits a non-linear stationary temperature profile $T_i = \langle p_i^2/2 \rangle_{ss} = \mathcal{T}(i/N)$, which was computed analytically for fixed and free boundary conditions --- surprisingly the temperature-profile was different for these cases \cite{Lepri2009,Cividini2017}. The evolution of the non-stationary temperature profile $\mathcal{T}(x,\tau)$ (where the $\tau=t/N^{3/2}$ is the rescaled time) to the NESS profile was also studied \cite{Lepri2010}, where by eliminating the fast variables it was shown that $\mathcal{T}(x,\tau)$ satisfies an energy continuity equation. From an analysis of this equation it was noted that  the evolution appears to be similar to the fractional diffusion equation. However, so far this has not been clearly established and in particular an explicit representation of the corresponding fractional evolution operator is not known.
	In this Letter, we look at a simpler model of anomalous transport in one dimension where we derive the corresponding fractional evolution equation for the temperature profile inside a finite domain and show explicitly how this evolution approaches to the appropriate fractional diffusion  operator in the infinite domain. 
	
	This model consists of a finite one dimensional lattice of $L$ sites where each site carries a `stretch' variable $\eta_i,~i=1,2,...,L$ under an onsite external potential 
	$V(\eta_i)=\eta_i^2/2$. The lattice is attached to two thermal reservoirs at temperatures $T_\ell$ and $T_r$ on the left and right ends, respectively and subjected to a volume conserving stochastic noise. The dynamics of this model has two parts: (a) the usual deterministic part plus the Langevin terms coming from the baths and (b) a stochastic exchange part where $\eta$s from any two neighboring sites, chosen at random, are exchanged at some rate $\gamma$. The dynamics is given by
	\begin{align}
	\begin{split}
	\frac{d\eta_i}{dt}= &~V'(\eta_{i+1})-V'(\eta_{i-1}) \\
	&+\delta_{i,1}\left(-\lambda V'(\eta_1) + \sqrt{2 \lambda T_\ell} \zeta_\ell(t) \right) \\
	&+\delta_{i,L}\left(-\lambda V'(\eta_L) + \sqrt{2 \lambda T_r} \zeta_r(t) \right)\\
	&+\text{stochastic~exchange~at~rate~}\gamma
	\end{split}
	\label{dynamics-eta}
	\end{align}
	with fixed boundary conditions (BCs) $\eta_0=\eta_{L+1}=0$. Here $\zeta_{\ell,r}(t)$ are mean zero and unit variance, independent Gaussian white noises. Note that, in contrast to the HCME case, this dynamics has two conserved quantities: the `volume' $\eta_i$ and the energy $V(\eta_i)$. This model was first introduced by Bernardin and Stoltz  in the closed system setup \cite{Bernardin2012} where starting from the harmonic chain with Hamiltonian given earlier, they have treated the positions $q_i$s and the momenta $p_i$s on the same footing. Note that for harmonic chain, the dynamics of the `stretch' variable $r_i=q_{i+1}-q_i$ and the momentum variable are similar: $\dot{r}_i= p_{i+1}-p_i$ and $\dot{p}_i=r_{i+1}-r_i$ for $i=1,2,...,N$. Hence for $N=L/2$, defining $\eta_{2j-1}=r_j$ and $\eta_{2j}=p_j$, one finds that both the above equations can be expressed in a single equation: $\dot \eta_m=\eta_{m+1}-\eta_{m-1}$ for $m=1,2,...,L$. The system can also be interpreted as a fluctuating interface where the algebraic volume of the interface at site $m$ is given by $\eta_m$ and the energy $V(\eta)=\eta^2/2$ \cite{Bernardin2012}. Hence, the stochastic exchange part in Eq.~\eqref{dynamics-eta} can be thought of as a volume-energy conserving noise. We call this model as `harmonic chain with volume exchange' (HCVE).
	
	It has been shown that the HCVE model defined on an isolated infinite one dimensional lattice (\emph{i.e.} $\lambda=0$ in Eq.~\eqref{dynamics-eta} with $i=-\infty,..,-1,0,1,..,\infty$) exhibits super diffusion of energy~\cite{Bernardin2016}: 
	\begin{align}
	\begin{split}
	&\partial_t e(x,t)= -  \mathbb{L}_{\infty}[e(x,t)],~\\
	&\mathbb{L}_{\infty}= \frac{1}{\sqrt{2\gamma}}[(-\Delta)^{3/4}-\nabla (-\Delta)^{1/4}],
	\end{split}
	\label{eta-frac}
	\end{align}
	where the skew-fractional operator $\mathbb{L}_{\infty}$ has the Fourier representation $|q|^{3/2}(1-\iu~\text{sgn}(q))$ with $\iu=\sqrt{-1}$ and sgn({\it q}) is the Signum function.
	In this paper, however, we consider the HCVE model on a finite lattice of size $L$ in open set up {\emph{i.e.}} connected to heat baths at the two ends as described in Eq.~\eqref{dynamics-eta}. It is known that in this case also, as in HCME, the stationary current scales as $j \sim L^{-1/2}$ \cite{Bernardin2016}. 
	
	{\it Results}.- We explicitly find that in the large $L$ limit the stationary `energy' current $j_{ss}$ is given, in the leading order, by
	\begin{align}
	j_{ss} = \frac{1}{2} \sqrt{\frac{\pi}{\gamma}}~\frac{(T_\ell-T_r)}{\sqrt{L}}.
	\label{j_ss}
	\end{align}
	In the non-stationary regime, we numerically find that the temperature profile $T_i(t)=\langle V(\eta_i(t))\rangle$ and the two-point correlations $C_{i,j}(t)=\langle \eta_i(t)\eta_j(t)\rangle$ for $i \neq j$ have the following scaling forms 
	\begin{align}
	\begin{split}
	T_i(t) &= \mathcal{T} \left(\frac{i}{L},\frac{t}{L^{3/2}} \right) \\
	C_{i,j}(t)&= \frac{1}{\sqrt{L}}\mathcal{C}\left(\frac{|i-j|}{\sqrt{L}},\frac{i+j}{2L}, \frac{t}{L^{3/2}}\right),
	\end{split}
	\label{scaling}
	\end{align}
	in the leading order for large $L$. The scaling functions $\mathcal{T}(y,\tau)$ and $\mathcal{C}(x,y,\tau)$ 
	satisfy the following equations inside the domain $\mathcal{D}=\{ 0\leq x \leq \infty~;~0 \leq y \leq~1\}$:
	\begin{align}
	\partial_y \mathcal{C}(x,y,\tau)&= - \gamma \partial_x^2  \mathcal{C}(x,y,\tau) 
	\label{evo-eq-1} \\
	\partial_y \mathcal{T}(y,\tau)&= - 2\gamma \left[ \partial_x  \mathcal{C}(x,y,\tau)\right]_{x=0} 
	\label{evo-eq-2} \\ 
	\partial_\tau \mathcal{T}(y,\tau) &= 2 \partial_y  \mathcal{C}(0,y,\tau),
	\label{evo-eq-3}
	\end{align}
	with $\mathcal{C}(x,y,0)|_{x \to \infty}=0$ and $\mathcal{C}(x,y,0)=0$. We find that the exact solutions of these equations are given by
	\begin{align}
	\mathcal{T}(y,\tau)&=\mathcal{T}_{ss}(y)+\mathcal{T}_r(1-y,\tau)  \label{T}\\
	\mathcal{C}(x,y,\tau)&=\mathcal{C}_{ss}(x,y)+\mathcal{C}_r(x,1-y,\tau). \label{C}
	\end{align}
	In the above equation, NESS part of the profiles are 
	\begin{align}
	\begin{split}
	\mathcal{T}_{ss}(y)&= T_r + (T_\ell-T_r)~\sqrt{1-y}, \\ 
	\mathcal{C}_{ss}(x,y)&= -\frac{T_\ell-T_r}{4} \sqrt{\frac{\pi}{\gamma}}~\text{erfc} \left( \frac{x}{\sqrt{4 \gamma(1-y)}}\right),
	\end{split}
	\label{ss-soln} 
	\end{align}
	and the relaxation parts are 
	\begin{flalign}
	&\mathcal{C}_r(x,z,\tau)=- \int_0^z \frac{\exp\left(-\frac{x^2}{4\gamma (z-z')} \right)}{\sqrt{4 \pi \gamma (z-z')}}~\frac{\partial \mathcal{T}_r(z',\tau)}{\partial z'}dz' \label{C-sol}
	\end{flalign}
	where $\mathcal{T}_r(z,\tau)$  satisfies the following continuity equation:
	\begin{equation}
	\partial_\tau \mathcal{T}_r(z,\tau)= \frac{1}{\sqrt{\pi \gamma}} \partial_z \left[ \int_0^z dz' ~\frac{\partial_{z'}\mathcal{T}_r(z',\tau)}{\sqrt{z-z'}}~ \right],
	\label{heat-ano}
	\end{equation}
	inside the domain $0 \le z \leq 1$ with BCs $\mathcal{T}_r(0,\tau)=\mathcal{T}_r(1,\tau)=0$. 
	The relaxation parts $\mathcal{T}_r(z,\tau)$ and $\mathcal{C}_r(x,z,\tau)$ describe the approach towards the NESS solutions in the $\tau \to \infty$ limit. The equations Eq.~\eqref{j_ss}, Eq.~\eqref{ss-soln}, Eq.~\eqref{C-sol} and Eq.~\eqref{heat-ano}, comprise our main results. Note that the evolution of the temperature in Eq.~\eqref{heat-ano} is indeed given by a linear but non-local equation defined inside a finite domain $0~\le~z~\le~1$. However, following a similar calculation for infinite system we later show that Eq.~\eqref{heat-ano} reduces to Eq.~\eqref{eta-frac} \redw{\cite{SM}}. This establishes, without ambiguity, that the non-local operator in Eq.~\eqref{heat-ano} is the correct finite domain representation of the fractional operator $\mathbb{L}_{\infty}$ in Eq.~\eqref{eta-frac}. Another point to note that the temperature profile in SS, $\mathcal{T}_{ss}(y)$, is asymmetric under space reversal as the microscopic model itself does not have such symmetry. As a result, any locally created perturbation splits into one traveling sound mode and one non-moving heat mode. This is in contrast to the HCME model where one observes two sound modes moving in opposite directions in addition to a non-moving heat mode \cite{Spohn2016, Bernardin2012}. Consequently, in this case, there is singularity in $\partial_y \mathcal{T}_{ss}(y)$ only at one boundary and we find that the meniscus exponent \cite{Lepri2011} is again $1/2$ as in the HCME model 
	with fixed boundary conditions. Interestingly, it turns out that for this boundary condition, both the temperature and the correlation become independent of the strength of coupling~$\lambda$ with the heat baths in the large $L$ limit.

	%
	
	\begin{figure}[t]
		\centering
		\includegraphics[scale=0.45]{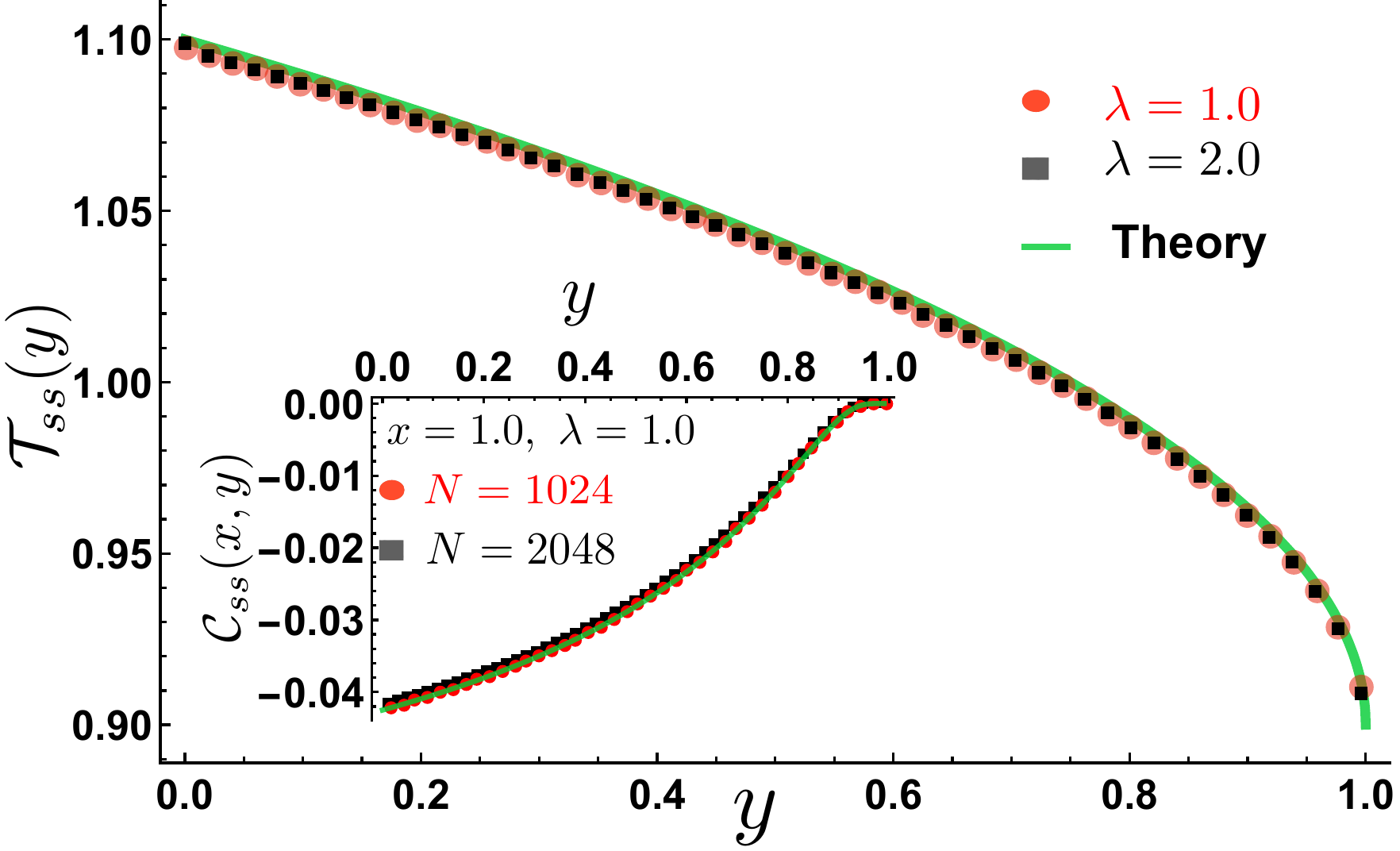}
		\caption{Numerical verification of the analytical NESS predictions for $\mathcal{T}_{ss}(y)$ and $\mathcal{C}_{ss}(x,y)$ in Eq.~\eqref{ss-soln}. Symbols are obtained from simulations with $\omega=\gamma=1,~T_\ell=1.1,~T_r=0.9$ and $N=1024$, whereas the solid lines are from theory.}
		\label{fig:ss-profiles}
	\end{figure}
	
	{\it Derivation of the results:}
	We start with the Fokker-Planck (FP) equation associated to the dynamics Eq.~\eqref{dynamics-eta}, which describes the evolution of  the joint distribution $P(\vec \eta,t)$ of $\vec \eta = (\eta_1,\eta_2,\cdots,\eta_L)$ at time $t$ :
	\begin{align}
	\partial_tP(\vec \eta ,t) &= [\mathcal{L}_{\ell}+\mathcal{L}_{b}+\mathcal{L}_{ex}]~P(\vec \eta ,t), \label{FP}
	\end{align}
	where, 
	$\mathcal{L}_\ell$ is the Liouvillian part, $\mathcal{L}_{b}$ contains the effects of the Langevin baths at the boundaries and $\mathcal{L}_{ex}$ represents the contribution from the exchange noise. Explicit expressions of these operators are given in \redw{\cite{SM}}. 
	While it is usually hard to compute explicitly the full joint distribution $P(\vec \eta, t)$,  computing various correlations and fluctuations is more tractable and they contain the most useful information about the non-equilibrium state. In this paper, for HCVE model we compute {\it all} the two point 
	correlations $\langle \eta_i(t) \eta_j(t) \rangle$ exactly in the limit of large system size.
	
	Starting from the FP equation in Eq.~\eqref{FP}, we obtain the dynamical equations satisfied by $T_{i}=\langle \eta_i^2(t) \rangle$ and $C_{i,j}=\langle \eta_i(t) \eta_j(t) \rangle$ for $i \neq j$ in the bulk:
	\begin{align}
	\begin{split}
	&\dot{C}_{ij}=C_{i+1,j}-C_{i-1,j}+C_{i,j+1}-C_{i,j-1}  \\
	&~~~~+\gamma[C_{i-1,j}+C_{i+1,j}+C_{i,j-1}+C_{i,j+1}-4C_{i,j}], \\
	&\dot{C}_{i,i+1}=T_{i+1}-C_{i-1,i+1}+C_{i,i+2}-T_{i} \\
	&~~~~~~~~~~~+\gamma[C_{i-1,i+1}+C_{i,i+2}-2C_{i,i+1}], \\
	&\dot{T}_i=2[C_{i,i+1}-C_{i-1,i}]+\gamma[T_{i+1}+T_{i-1}-2T_i].
	\end{split}
	\label{corr-eq}
	\end{align}
	Rest of the equations at the boundaries are given in \redw{\cite{SM}}. 
	Fortunately the equations for two point correlations do not involve higher order correlations, this allows us to solve these equations  analytically, in the $L \to \infty$ limit. 
	
	To proceed we follow the strategy in \cite{Lepri2010}. We first solve these equations numerically to observe that, for large $L$ the solutions have the scaling properties as given in Eq.~\eqref{scaling} where we have two length scales of $O(L)$ along the diagonal ($i+j=$constant) and of $O(\sqrt{L})$ along perpendicular to the diagonal ($|i-j|=$constant) direction, and a time scale of $O(L^{3/2})$. This time scale can be anticipated from the propagator $e^{-|q|^{3/2}[1-\iu \text{sgn}(q)]t}$ of the Eq.~\eqref{eta-frac} in Fourier space. The two length scales are understood by looking at the orders of the $C_{i,j}$ and $T_i$, and their derivatives numerically \redw{\cite{SM}}.
	These observations suggest that we look for solutions of Eq.~\eqref{corr-eq} in the scaling form Eq.~\eqref{scaling}. Inserting these forms in Eq.~\eqref{corr-eq}, and expanding in $\epsilon=L^{-1/2}$, we obtain three coupled linear differential equations as given in~Eq.~\eqref{evo-eq-1}-Eq.~\eqref{evo-eq-3} in the leading order \redw{\cite{SM}}. Interestingly, the scaled correlation function $\mathcal{C}$ relaxes very fast over much shorter time scale [$O(L)$] compared to the evolution time scale  [$O(L^{3/2})$] of  the temperature field $\mathcal{T}$. Due to this fact, Eq.~\eqref{evo-eq-1} and Eq.~\eqref{evo-eq-2} do not involve the time derivative.  As a result the correlation function $\mathcal{C}$ evolves adiabatically obeying the (anti-)diffusion Eq.~\eqref{evo-eq-1}, with a drive at the boundary by the time dependent temperature field through Eq.~\eqref{evo-eq-2}. The equation for the temperature profile given in Eq.~\eqref{evo-eq-3} is in the expected continuity equation.

	In the NESS the equations \eqref{evo-eq-1}-\eqref{evo-eq-3} become simpler since $\partial_\tau \mathcal{T} \to 0$ as $\tau \to \infty$ implying $\mathcal{C}_{ss}(0,y)=d$. 
	Now making the variable transformation $z=(1-y)$, the problem of finding $\mathcal{C}_{ss}$ reduces to solving a diffusion equation with its value at $x=0$ held fixed for all $y$. It is easy to show that the solution is given by $\mathcal{C}_{ss}(x,y) = d~\text{erfc}(x/\sqrt{4\gamma (1-y)})$ where 
	$\text{erfc}(v)$ is the complementary error function  \redw{\cite{SM}}.
	Now inserting this solution in Eq.~\eqref{evo-eq-2} and solving with boundary conditions 
	$\mathcal{T}_{ss}(0)=T_\ell$ and $\mathcal{T}_{ss}(1)=T_r$, we get the explicit expression 
	Eq.~\eqref{ss-soln}. 
	In Fig. \ref{fig:ss-profiles} we verify the analytical results for $T_{ss}$ and $\mathcal{C}_{ss}$ numerically, where we observe nice agreement. One can easily identify the microscopic current from the equation for $\dot{T}_i$ in Eq. \eqref{corr-eq} as  $j = -2C_{i,i+1} -\gamma(T_{i+1}-T_i)$ which in the steady state for large $L$ provides
	$j_{ss}=-2\mathcal{C}_{ss}(0,y)/\sqrt{L}$. Note that the term $-\gamma(T_{i+1}-T_i)$ contributes at $\mathcal{O}(1/L)$. Now inserting the expression of $\mathcal{C}_{ss}(0,y)$ from Eq. \eqref{ss-soln}, one obtains the expression for $j_{ss}$ given in Eq.~\eqref{j_ss}.
	
	\begin{figure}[t]
		\centering
		\includegraphics[scale=0.45]{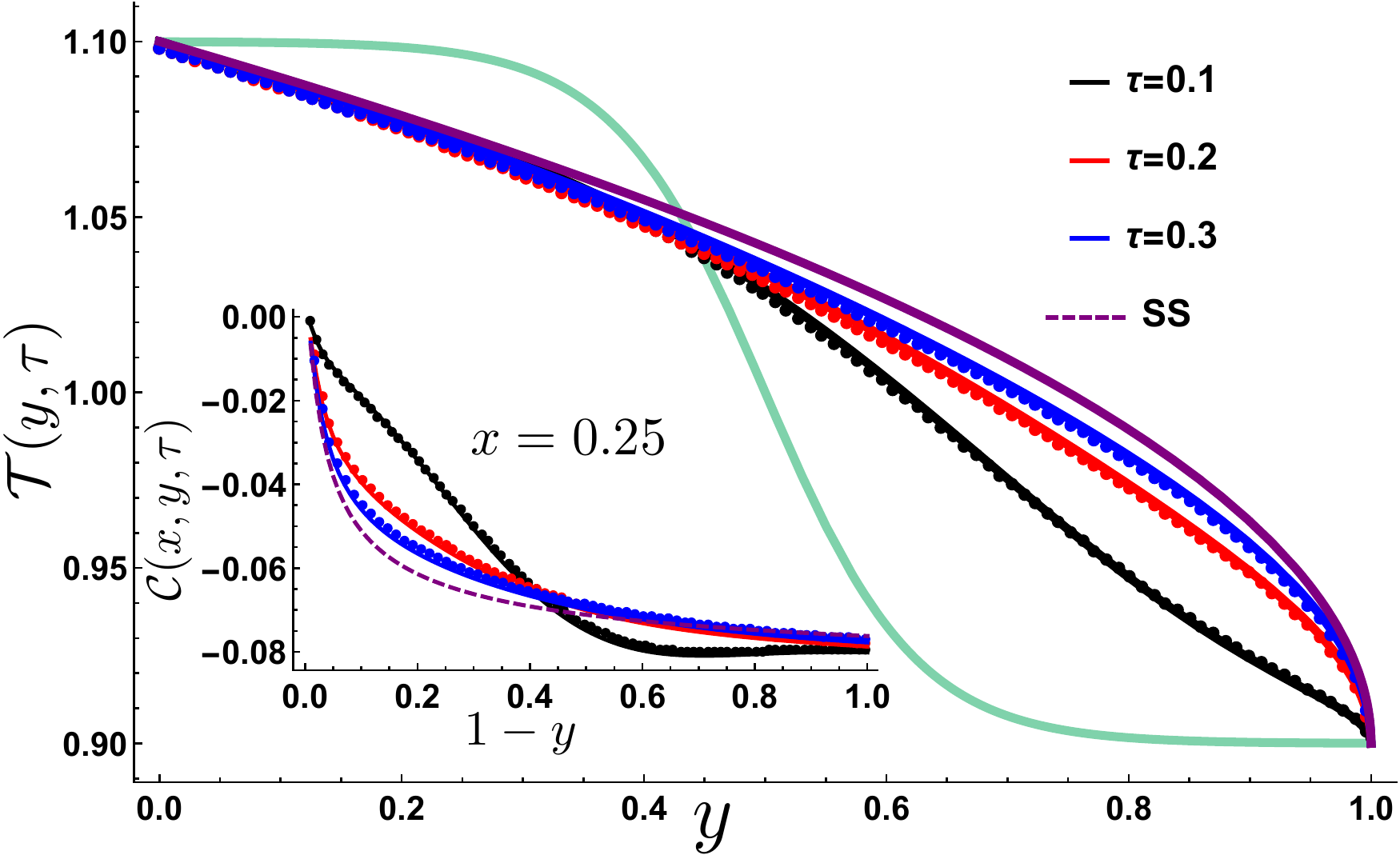}
		\caption{Numerical verification of the evolution of the temperature profiles $\mathcal{T}(y,\tau)=\mathcal{T}_{ss}(y)+\mathcal{T}_r(1-y,\tau)$ obtained using the solution of Eq.~\eqref{heat-ano}. Inset shows the verification of the  correlation $\mathcal{C}(x,y,\tau)=\mathcal{C}_{ss}(x,y)+\mathcal{C}_r(x,1-y,\tau)$ given in Eq.~\eqref{C} where $\mathcal{C}_r$ is computed using the solution in Eq.~\eqref{C-sol}. The green line and the dashed purple line represent the initial and the NESS temperature profiles, respectively. Symbols are obtained from simulations with $\lambda=\gamma=1,~T_\ell=1.1,~T_r=0.9$ and $L=2048$, and the solid lines are from theory.}
		\label{fig:evolution}
	\end{figure}
	
	We now focus on the relaxation to the NESS. It is often convenient to separate the relaxation part as done in Eq.~\eqref{T} and Eq.~\eqref{C} where $\mathcal{T}_r(z,\tau)$ and $\mathcal{C}_r(x,z,\tau)$ describes the approach towards the NESS solutions in Eq.~\eqref{ss-soln}. It is easy to see that $\mathcal{C}_r(x,z,\tau)$ and 
	$ \mathcal{T}_r(z,\tau)$ satisfies the following equations 
	\begin{align}
	\partial_z \mathcal{C}_r(x,z,\tau)&= \gamma \partial_x^2  \mathcal{C}_r(x,z,\tau), 
	\label{evo-eq-1-diff} \\
	\partial_z \mathcal{T}_r(z,\tau)&= 2\gamma \left[ \partial_x  \mathcal{C}_r(x,z,\tau)\right]_{x=0}, \label{BC-C} \\
	\partial_\tau \mathcal{T}_r(z,\tau) &= -2 \partial_z  \mathcal{C}_r(0,z,\tau), \label{T-C-eq}
	\end{align}
	with initial condition $\mathcal{C}_r(x,z,0)=0$ and BC $ \mathcal{C}_r(x,z,\tau)|_{x \to \infty}=0$. The above equations are obtained from Eq.~\eqref{evo-eq-1}-\eqref{evo-eq-3} after subtracting the steady state part and then making the variable transformation $z=(1-y)$.
	Note that the BC in Eq.~\eqref{BC-C} acts like a current source, at $x=0$ boundary, to the diffusion Eq.~\eqref{evo-eq-1-diff}. It is easy to show \redw{\cite{SM}} that the solution of this equation is given precisely by Eq.~\eqref{C-sol}. Now, inserting this solution in Eq.~\eqref{T-C-eq}
	and performing some simplifications we arrive at the non-local evolution Eq.~\eqref{heat-ano} (see \redw{\cite{SM}}) with boundary conditions $\mathcal{T}_r(0,\tau)=\mathcal{T}_r(1,\tau)=0$. Hence it is natural to expand the solution in $\sin$ basis $\alpha_n(z)=\sqrt{2}~\sin(n \pi z)$, $n=1,2,...$: 
	$\mathcal{T}_r(z,\tau)=\sum_{n=1}^\infty \hat{\mathcal{T}}_r(n,\tau)~\alpha_n(z)$
	for all $\tau \geq 0$. Inserting this form in Eq.~\eqref{heat-ano} and simplifying we find that the coefficients $\hat{\mathcal{T}}_r(n,\tau)$ satisfy the following matrix equation 
	\begin{align}
	\frac{d \hat{\mathcal{T}}_r}{d \tau} = \kappa {\bf B}~\hat{\mathcal{T}}_r, \label{T-mat-inf}
	\end{align}
	where ${\bf B}$ is an infinite order matrix 
	with elements ${\bf B}_{m,n}$ specified in \redw{\cite{SM}} and $\kappa =  1/\sqrt{ \pi \gamma} $. While it is difficult to solve this infinite order matrix equation analytically,  we solve it numerically by truncating it at some finite order. In Fig. \ref{fig:evolution}, we compare the evolution from this numerical solution with the same obtained from direct numerical simulation of Eq.~\eqref{dynamics-eta} and observe nice agreement. Using this solution in Eq.~\eqref{C-sol} we obtain $C(x,z,\tau)$ in Eq.~\eqref{C} which we also compare with simulation results in the inset of Fig. \ref{fig:evolution} and again observe good agreement.
	
	From the numerical solution of Eq.~\eqref{T-mat-inf} we, in addition, observe that the eigenvalues obtained , are in general complex and are of the form $\lambda_n \approx |n\pi|^{3/2}(1-\iu \text{sgn})$, for large $n >0$, while at small values of $n$ the eigenvalues deviate from this behaviour \redw{\cite{SM}}. We confirm that this is not due to the the truncation of the matrix but an artefact of the finiteness of the system. Note that the large $n$ behavior of $\lambda_n$ is similar to the Fourier spectrum of the non-local operator $\mathbb{L}_\infty$ in Eq.~\eqref{eta-frac} describing the evolution in infinite system. Hence it is interesting to see if one recovers the evolution Eq.~\eqref{eta-frac} in the infinite system limit. One can follow the same calculation procedure on an infinite lattice as presented before for finite lattice and arrive at an evolution equation for temperature profile similar to Eq.~\eqref{heat-ano} with only difference being that the lower limit of the integral on the rhs is $-\infty$, as the equation is now valid in $-\infty \le y \le \infty$. Now taking Fourier transform on both sides it is quite easy to show that the evolution equation in infinite space indeed reduces to the skew-fractional equation in Eq.~\eqref{eta-frac}. A more direct and detailed proof is given in \redw{\cite{SM}}.

	
	{\it Conclusion}: In this Letter, we have studied anomalous transport in a one-dimensional system with two conserved quantities, in the open system setup. Starting from a microscopic description and acquiring knowledge about scaling properties from numerical studies, we derive exact expressions of the temperature profiles and the two point correlations in the steady state.  We also study the evolution of these quantities towards steady state. 
	We explicitly show that the evolution of the temperature profiles in this model is governed by a non-local operator defined inside a finite domain which correctly takes the previously obtained infinite system representation. We provide numerical verifications of the analytical results.
	Our work provides the first clear and transparent microscopic derivation of non-local heat equation describing anomalous transport in finite geometry and its connection to the corresponding skew-fractional equation in the infinite domain.

	We thank C. Bernardin, C. Mej\'ia-Monasterio, S. Olla and  S. Lepri for very useful discussions. AK 
	acknowledges support from DST grant under project No. ECR/2017/000634.
	%
	%
	\vspace{-0.2cm}
	\bibliographystyle{unsrt}
	\bibliography{BSmodel}
	\nopagebreak



\pagebreak
\widetext
\begin{center}
\textbf{\large Supplementary Material: Anomalous heat equation in a system connected to thermal reservoirs}
\\{Priyanka$^1$, Aritra Kundu$^1$, Abhishek Dhar$^1$, Anupam Kundu$^1$\\	$^1$International Centre for Theoretical Sciences, Tata Institute of Fundamental Research,\\ Bengaluru 560089, India}

\begin{center}{\textbf{Abstract}}
	{
	
		We here provide the details of calculations for the results presented in the main text.
	}
\end{center}

\end{center}
\setcounter{equation}{0}
\setcounter{figure}{0}
\setcounter{table}{0}
\setcounter{page}{1}
\makeatletter
\renewcommand{\theequation}{S\arabic{equation}}
\renewcommand{\thefigure}{S\arabic{figure}}
\renewcommand{\bibnumfmt}[1]{[S#1]}
\renewcommand{\citenumfont}[1]{S#1}
\section{Dynamical operators and Discrete equations}
The explicit expression of the deterministic part of the Fokker Planck (FP) equation given
by operators ${\cal{L}}_l$, ${\cal{L}}_b$, defined as
\eqa{
	{\cal L}_l &=\sum_{i=1}^{L-1}(V'(\eta_{i+1})-V'(\eta_{i-1}))\partial_{\eta_i}-V'(\eta_{L-1})
	\partial_{\eta_L}+V'(\eta_2)\partial_{\eta_1} \nn \\
	{\cal L}_b &= \lambda T_\ell~\partial^2_{\eta_1}-\lambda  \partial_{\eta_1} V'(\eta_1)
	+\lambda T_r~\partial^2_{\eta_L}-\lambda  \partial_{\eta_L} V'(\eta_{L}), \nn
}
where $T_\ell$ and $T_r$ are the temperatures of the reservoirs on the left and right, respectively.
The stochastic part ${\cal{L}}_{ex}$ is given as
\eqa{
	{\cal L}_{ex}=\gamma \left( \sum_{i=1}^{L-1}P(\vec{\eta}_{i,i+1})-P(\vec{\eta} )\right)
}
where $\vec{\eta}_{i,i+1}$ denote the configuration after the exchange of variable $i$ with $i+1$.
The two point function defined as, $C_{i,j}=\langle \eta_i\eta_j\rangle$. From the above FP equation for $V(\eta)=\eta^2/2$, the dynamical equation $C_{i,j}=\langle \eta_i\eta_j\rangle$s can easily be written as 
\eqa{
	\dot{C}_{ij}~&=C_{i+1,j}-C_{i-1,j}+C_{i,j+1}-C_{i,j-1} 
	+\gamma[C_{i-1,j}+C_{i+1,j}+C_{i,j-1}+C_{i,j+1}-4C_{i,j}], ~~1 < i,j < L \\
	\dot{C}_{i,i+1}~&=T_{i+1}-C_{i-1,i+1}+C_{i,i+2}-T_{i} 
	+\gamma[C_{i-1,i+1}+C_{i,i+2}-2C_{i,i+1}], ~~1 < i < L  \\
	\dot{T}_i~&=2[C_{i,i+1}-C_{i-1,i}]+\gamma[T_{i+1}+T_{i-1}-2T_i],~~1 \leq i \leq  L
	\label{eqs:corr-eq}
}
The dynamical equations at the boundaries are given by 
\begin{itemize}
	\item[$1.$] for $i=j=1$
	\eqa{
		\dot{T}_{1}=2\lambda T_\ell+2C_{1,2}-2\lambda T_{1}+\gamma[T_{2}-T_{1}]
	}
	\item[$2.$] for $i=j=L$
	\eqa{
		\dot{T}_{L}=2 \lambda T_r-2C_{L-1,L}-2\lambda T_L+\gamma[T_{L-1}-T_L]
	}
	\item[$3.$] $i=1$ and $1<j<L$
	\eqa{
		\dot{C}_{1,j}=C_{2,j}-\lambda C_{1,j}+C_{1,j+1}-C_{1,j-1}+\gamma[ C_{1,j-1}+C_{1,j+1}+C_{2,j}-3C_{1,j}]
	}
	\item[$4.$] $j=L$ and $1<i<L$
	\eqa{
		\dot{C}_{i,L}=C_{i+1,L}-C_{i-1,L}-C_{i,L-1}-\lambda C_{i,L}+\gamma[C_{i-1,L}+C_{i+1,L}+C_{i,L-1}-3C_{i,L}]
	}
	\item[$5.$] $i=1$ and $j=L$
	\eqa{
		\dot{C}_{1,L}=C_{2,L}-C_{1,L-1}-2\lambda C_{1,L}+\gamma[C_{2,L}+C_{1,L-1}-2C_{1,L}]
	}
\end{itemize}
\begin{figure}[t!]
	\begin{center}
		\includegraphics[scale=0.8]{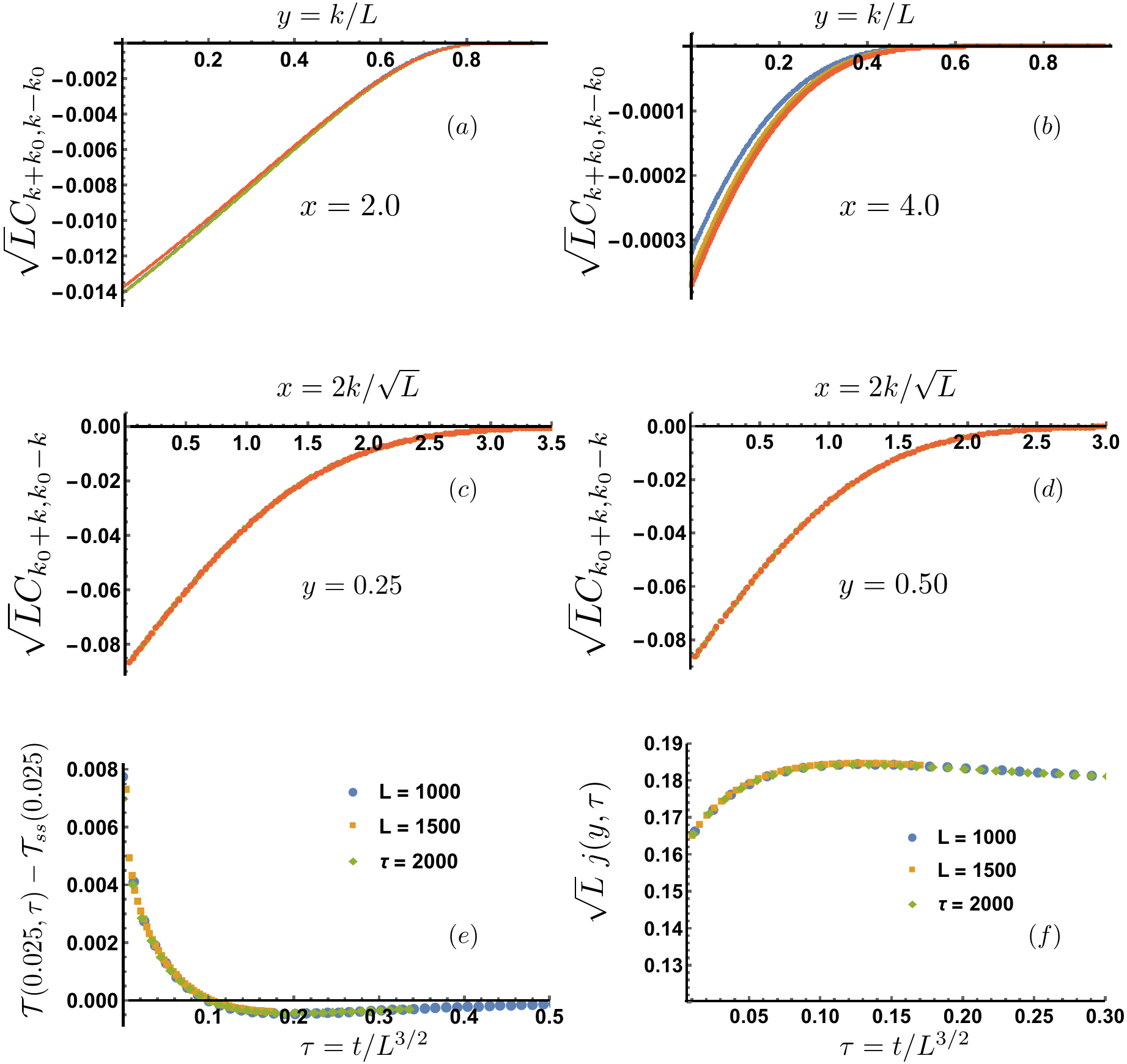}
	\end{center}
	\caption{Data collapse of the correlation functions and temperature profile confirming the scaling behaviors in Eqs.~\eqref{eqs:sc-C} and \eqref{eqs:sc-T}. 
		Figures (a) and (b) show the data collapse as a function of the scaling variable $y=(i+j)/2L$ with four systems sizes $L=1000$ (Blue), $L=2000$ (Orange), $L=3000$ (Green) and $L=4000$ (Red), for two fixed values of $x=|j-i|/\sqrt{L}$. Figures (c) and (d), show the data collapse as a function of the scaling variable $x$  with the above four system sizes for two fixed values of $y$. The collapse are so good that other colors not visible.  Figure (e) describes the scaling behavior for the evolution of the temperature $\mathcal{T}(y,\tau)=T_{\lfloor yL \rfloor}(\tau L^{3/2})$ at a fixed position $y=0.025$ for different system sizes as a function of the scaled time $\tau=t/L^{3/2}$. Note that, the temperatures are of $\mathcal{O}(1)$ whereas the correlations are of  $\mathcal{O}(1/\sqrt{L})$. Also note from figures (c) and (b) that $\mc{C}_{ss}(x \to \infty,z)=0$. Last figure (d) establishes that the current in the system is of order $1/\sqrt{L}$ and also evolves in scaled time  $\tau=t/L^{3/2}$. The other parameters in the simulation are $\gamma = \Lambda =1$, $T_\ell = 1.1, ~ T_r = 0.9 $.
	}
	\label{scaling_fig}
\end{figure}
\section{Derivation of continuum equations from discrete equations}
Now, we want to get a continuum description of the bulk discrete dynamical equations derived in the previous section. In the non-stationary regime, we numerically find that the temperature profile $T_i(t)=\langle \eta_i(t)^2\rangle$ and the two-point correlations $C_{i,j}(t)=\langle \eta_i(t)\eta_j(t)\rangle$ for $i \neq j$ have the following scaling forms 
\begin{align}
C_{i,j}(t)&= \frac{1}{\sqrt{L}}\mathcal{C}\left(\frac{|i-j|}{\sqrt{L}},\frac{i+j}{2L}, \frac{t}{L^{3/2}}\right), \label{eqs:sc-C} \\
T_i(t) &= \mathcal{T} \left(\frac{i}{L},\frac{t}{L^{3/2}} \right), \label{eqs:sc-T}
\end{align}
in the leading order for large $L$ which are also supported by numerical evidence shown in Fig~\ref{scaling_fig}. In figs.~(\ref{scaling_fig}a),~(\ref{scaling_fig}b, ~(\ref{scaling_fig}c) and ~(\ref{scaling_fig}d), verify the scaling behaviors of the correlations in Eq.~\eqref{eqs:sc-C}. Figures~(\ref{scaling_fig}c) and ~(\ref{scaling_fig}d), describes scaling behavior with respect to time. 
Using these, we define continuum ordinates as, $\frac{|i-j|}{\sqrt{L}}=x$, $\frac{|i+j|}{2L}=y$, $\frac{t}{L^{3/2}}=\tau$ and $\frac{1}{\sqrt{L}}=\epsilon$, where $x \in (0,\infty)$ and $y \in (0,1)$. In the following, we insert this scaling form and Taylor expand in $\epsilon = 1/\sqrt{L}$. Keeping terms to leading order in $\epsilon$ we obtain the continuum equations. 
%
%
\begin{enumerate}
	
	\item[1] \textbf{Bulk Equations, $|i-j| \ge 2 $}\\
	We start with the discrete equation in bulk:
	\eqa{
		\dot{C}_{i,j}=-\left( C_{i-1,j}-C_{i+1,j}+C_{i,j-1}-C_{i,j+1}-\gamma[C_{i,j-1}+C_{i,j+1}+C_{i+1,j}+C_{i-1,j}-4 C_{i,j}] \right),
	}
	using above scaling definitions, we can
	write the above mentioned discrete equation as,
\eqa{
		\epsilon^4 \p_\tau \mc{C}(x,y,\tau) = -\epsilon \left( {\cal C}(x-\epsilon,y-\frac{\epsilon^2}{2}) - {\cal C}(x+\epsilon,y+\frac{\epsilon^2}{2}) +
		{\cal C}(x+\epsilon,y-\frac{\epsilon^2}{2}) - {\cal C}(x-\epsilon,y+\frac{\epsilon^2}{2}) \nn \right. \\
		\left. 	-\gamma\left[{\cal C}(x+\epsilon,y-\frac{\epsilon^2}{2}) + {\cal C}(x-\epsilon,y+\frac{\epsilon^2}{2}) +
		{\cal C}(x+\epsilon,y+\frac{\epsilon^2}{2}) + {\cal C}(x-\epsilon,y-\frac{\epsilon^2}{2})-4{\cal C}(x,y)\right]\right) ~,
}
	which by Taylor expansion of each terms in $x,~y$ and $\tau$, we obtain the leading order terms for continuum dynamical equation as
	\eqa{
		\epsilon^4 \p_\tau \mc{C}(x,y,\tau)= 2\epsilon^3	 \p_y \mc{C}(x,y,\tau)+
		2 \gamma \epsilon^3	\p^2_x \mc{C}(x,y,\tau)~.
	}
	At the dominant order ($o(\epsilon^3)$), we find,
	\eqa{
		\p_y \mc{C}(x,y,\tau)+
		\gamma \p^2_x \mc{C}(x,y,\tau)=0
	}
	
	\item[2] \textbf{{Nearest neighbor term}, $j = i+1$}\\
	%
	%
	\eqa{
	\dot{C}_{i,i+1}= T_{i+1} - C_{i-1,i+1} + C_{i,i+2} - T_i + \gamma \left[ C_{i-1,i+1} + C_{i,i+2} - 2 C_{i,i+1}\right]
}
	after proper scaling, we get,
	\eqa{
	\epsilon^4 \p_\tau \mc{C}(\epsilon,y+\frac{\epsilon^2}{2},\tau) ={\cal T}(y+\epsilon^2) - {\cal T}(y) - \epsilon {\cal C}(2\epsilon,y+\epsilon^2)
	+\epsilon {\cal C}(2\epsilon,y)+ \gamma\epsilon \left[ {\cal C}(2\epsilon,y)+{\cal C}(2\epsilon,y+\epsilon^2)-2{\cal C}(\epsilon,y+\frac{\epsilon^2}{2})\right] ~.
}
	Expanding above equation in $x$ and $y$, and keeping the relevant order terms in $\epsilon$
	we get the continuum equation as
	\eqa{
	\epsilon^4\p_\tau \mc{C}(0,y,\tau) 
	=\epsilon^2\left(\p_y \mc{T}(y,\tau)+ 2\gamma \p_x \mc{C}(0,y,\tau) \right) +O(\epsilon^3)
}
	and hence to the dominating order, the governing continuum equation is
\eqa{
	\p_y \mc{T}(y,\tau)+ 2\gamma \p_x \mc{C}(0,y,\tau)=0
}
	
	\item[3] {\textbf{Diagonal term} $i=j$}\\
	Next is the diagonal term where $i=j$,
	\eqa{\dot{T}_i~&=2[C_{i,i+1}-C_{i-1,i}]+\gamma[T_{i+1}+T_{i-1}-2T_i].}
	this in continuum limit given as
	\eqa{
		\epsilon^3\p_\tau \mc{T}(y,\tau)= 2\epsilon \left[ \mc{C} (\epsilon,y+\frac{\epsilon^2}{2})- \mc{C} (\epsilon,y-\frac{\epsilon^2}{2}) \right]+ \gamma \left[ \mc{T}(y+\epsilon^2) +\mc{T}(y-\epsilon^2) -2\mc{T}(y)  \right]
	}
	After expansion, we arrive at
	\eqa{
		\epsilon^3\p_\tau \mc{T}(y,\tau) = 2\epsilon\left[ 
		\epsilon^2 \p_y\mc{C}(0,y,\tau)  +  \epsilon^3\frac{\gamma}{4}\p_y^2 \mc{T}(y,\tau) \right]+O(\epsilon^4)
	}
	Hence, the leading order term is
	\eqa{
		\p_\tau \mc{T} (y,\tau) = 2 \p_y \mc{C} (0,y,\tau)
	}
	
	\item[4] \textbf{Current}\\
	The microscopic energy current in the system is defined through
	\eqa{\p_t \la \eta_i^2 \ra = -[j_{i\to i+1} - j_{i-1 \to i}],}where $j_{i\to i+1}  = -2 C_{i,i+1} - \gamma (T_{i+1} - T_i)$. The stochastic part of the current decays as $O(1/L)$ and in  the macroscopic limit goes to zero. In the continuum limit, the deterministic part contributes in the leading order to provide, $j= -2 \mc{C} (0,y,\tau)/\sqrt{L}. $ 
\end{enumerate}

\section{Solution in the steady state as well as in the relaxation regime}
\noindent
The above analysis gives us the following bulk equations for the system,
\eqa{
	\p_y \mc{C}(x,y,\tau) &=
	-\gamma\p_{x}^2 \mc{C}(x,y,\tau)~,\label{eq:CyCx2-0}\\
	\p_y \mc{T}(y,\tau) &=-2\gamma \p_{x} \mc{C}(x,y,\tau)_{x=0}~\label{eq:TyC0x-0}\\
	\p_\tau \mc{T} (y,\tau) &= 2 \p_y \mc{C} (0,y,\tau) 
	\label{eqs:TtC0y-0}
}
Solutions of these equations for $\mc{C}(x,y,\tau)$ and $\mc{T}(y,\tau)$ have two parts. One part describes the steady state and the other part describing the relaxation to the steady state:
\begin{align}
\mathcal{T}(y,\tau)&=\mathcal{T}_{ss}(y)+\mathcal{T}_r(1-y,\tau)  \label{eqs:T}\\
\mathcal{C}(x,y,\tau)&=\mathcal{C}_{ss}(x,y)+\mathcal{C}_r(x,1-y,\tau). \label{eqs:C}
\end{align}
It is easy to show that $\mc{C}_{ss}(x,z)$ and $\mc{T}_{ss}(z)$  satisfy
\eqa{
	\p_z \mc{C}_{ss}(x,z) &=
	\gamma\p_{x}^2 \mc{C}_{ss}(x,z)~,\label{eqs:C_ss}\\
	\p_z {\cal T}_{ss}(z) &=2\gamma \p_{x} \mc{C}_{ss}(x,z)_{x=0}. \label{eqs:T_ss}
}
under the transformation of $z=1-y$ with boundary conditions $ \mc{C}_{ss}(x \to \infty,z) =0,~ \mc{T}_{ss}(0)=T_\ell$ and $\mc{T}_{ss}(1)=T_r$. On the other hand, the relaxation parts satisfy
\eqa{
	\p_z \mc{C}_r(x,z,\tau) &=
	\gamma\p_{x}^2 \mc{C}_r(x,z,\tau)~,\label{eqs:CyCx2}\\
	\p_z \mc{T}_r(z,\tau) &=2\gamma \p_{x} \mc{C}_r(x,z,\tau)_{x=0}~\label{eqs:TyC0x}\\
	\p_\tau \mc{T}_r(y,\tau) &= -2 \p_z \mc{C}_r(0,z,\tau). 
	\label{eqs:TtC0y}
}
with appropriate boundary conditions specified in Sec.\ref{ss:relationdynm}.

\subsection{Stationary state solution}
\noindent
In stationary state, we first solve Eq. \eqref{eqs:C_ss} and Eq. \eqref{eqs:T_ss}.  We want to solve these equations along with the boundary conditions  \eqa{(i)~\mc{C}_{ss} (x,z \to 0)=0,~~(ii)~\mc{C}_{ss}(x \to \infty,z)=0,~~(iii)~\mc{C}_{ss}(x=0,z) = d.} The initial condition in (i) is easy to understand. The boundary condition (ii) is obtained from our numerical observation [see fig.~\ref{scaling_fig}(c,d) ].
Where the last boundary condition (iii) is obtained by observing that LHS of \eqref{eqs:TtC0y} is zero in steady state; hence ($\p_z \mc{C}_{ss} (0,z)= 0$). The unknown constant $d$ will be fixed by the temperatures at the boundary. The first equation is easy to solve by taking Laplace transform in $z$ along with boundary conditions. Finally inverting the Laplace transform, we find, the solution is given by,
\eqa{
	\mc{C}_{ss} (x,z)&=d~ \text{erfc}\left[\frac{x}{\sqrt{4\gamma z}}\right]
}
where, $\text{erfc}$ is the complimentary error function defined as $\text{erfc}(x) = 1-\frac{2}{\sqrt{\pi }}\int _{0}^{x}d t e^{-t^2}$.
Now, using this solution in the Eq.~\ref{eqs:T_ss}, we get
\eqa{
	\p_z \mc{T}_{ss}(z) =& -d~\sqrt{\frac{4\gamma}{\pi z}} \nn \\
	%
	%
	{\cal T}_{ss}(z)=&{\cal T}_{ss}(0)-2d~\sqrt{\frac{4\gamma z}{\pi }} \label{eqs:sstemp1}
}
The constants ${\cal T}_{ss}(0)$ and $d$ will now be determined from the boundary conditions of temperature field, $\mc{T}(z=0) = T_r$ and $\mc{T}(z=1) = T_\ell$,. We finally have,
\eqa{
	T_\ell-T_r=&~-2 d ~\sqrt{\frac{4\gamma}{\pi }}~\nn\\
	d~=&~-\frac{\Delta T}{4}\sqrt{\frac{\pi}{\gamma}}~,
}
where $\Delta T=(T_\ell-T_r)$ is the temperature difference between the left and right heat baths.  Reverting now back to $y$ variables using $z=1-y$, the exact expressions for the steady state temperature profile and correlations are,
\eqa{
	\mc{ T}_{ss}(y)=&T_r + \Delta T \sqrt{1-y},\\
	\mc{C}_{ss}(x,y) =& -\frac{\Delta T}{4}\sqrt{\frac{\pi}{\gamma}}\text{erfc}\left[\frac{x}{\sqrt{4\gamma (1-y)}}\right]\\
}
Hence the current in the system is given by $j_{ss} = \frac{\mc{J}_{ss}}{\sqrt{L}}$, where,
\eqa{j_{ss}=& -\frac{2\mc{C}_{ss}(0,y)}{\sqrt{L}} = \frac{ \Delta T}{2} \sqrt{\frac{\pi}{\gamma}}~\frac{1}{\sqrt{L}}}


\subsection{Solution in the relaxation regime}\label{ss:relationdynm}
\noindent
Here we solve Eqs. \eqref{eqs:CyCx2},  \eqref{eqs:TyC0x} and \eqref{eqs:TtC0y} (Note that these are also 
Eqs.~(16),(17) and (18) of the main text). For convenience, we rewrite these equations here,
\begin{align}
\partial_z \mathcal{C}_r(x,z,\tau)&= \gamma \partial_x^2  \mathcal{C}_r(x,z,\tau), 
\label{eqs:evo-eq-1-diff} \\
\partial_z \mathcal{T}_r(z,\tau)&= 2\gamma \left[ \partial_x  \mathcal{C}_r(x,z,\tau)\right]_{x=0}, \label{eqs:BC-C} \\
\partial_\tau \mathcal{T}_r(z,\tau) &= -2 \partial_z  \mathcal{C}_r(0,z,\tau). \label{eqs:T-C-eq}
\end{align}
We want to solve \eqref{eqs:evo-eq-1-diff} with boundary conditions $\mathcal{C}_r(x,z) |_{x \to \infty} = 0$ and \eqref{eqs:BC-C}. The Greens function $g(x,z)$ of this equation with above BC's   satisfies, $\partial_z g(x,z)= \frac{\gamma}{2} \partial_x^2  g(x,z)$, where, $g(x,z)$ is given by $g(x,z) = \sqrt{4 \gamma z} h(x/\sqrt{4 \gamma z})$ where, $h(w) = \frac{e^{-w^2}}{\pi} -w~ \text{erfc}(w)$, hence, the general time dependent solution is written as,
\begin{eqnarray}
\mc{C}_r(x,z,\tau) =2 \int_0^z dx' \frac{e^{ -(x-x')^2/(4 \gamma z)} }{\sqrt{4 \pi \gamma z}}\mc{C}_r(x',0,\tau) &-  \frac{1}{2\gamma} \left(\int_0^z dz' (g(x,z-z') \p_{z'}^2 \mc{T}_r(z',\tau)) \right) \nonumber \\
&~~-  \frac{1}{2\gamma}~\p_{z'} \mc{T}_r(z',\tau) g(x,z)|_{z' \to 0} \label{eqs:cxygen}
\end{eqnarray}
With the initial condition $\mc{C}_r(x,0,\tau) = 0$ the first term drops out. It is easy to check that the remaining part satisfies \eqref{evo-eq-1-diff} with boundary condition \eqref{BC-C} as follows, 
\eqa{\p_x \mc{C}_r(x,z,\tau)|_{x \to 0} =&~-  \frac{1}{2\gamma}\left( \int_0^z dz' \p_x g(x,z-z') ) \p_{z'}^2 \mc{T}_r(z',\tau) + \p_x g(x,z) ) \p_{z'} \mc{T}_r(z',\tau) |_{z' \to 0}\right)  \nn \\
	=&\frac{1}{2\gamma}\left( \int_0^z dz' \p_{z'}^2 \mc{T}_r(z',\tau) + \p_{z'} \mc{T}_r(z',\tau) |_{z' \to 0} \right)
	=~  \frac{1}{2\gamma} \p_{z} \mc{T}_r(z,\tau) 
}
where we have used $\p_x g(x,z) |_{x \to 0} = -1$.
Further using the fact that $g(x,z-z') \p_{z'}T_r(z')|_{z' \to z} \to  0$ we can simplify \eqref{eqs:cxygen} as,
\eqa{\mc{C}_r(x,z,\tau) = &-  \frac{1}{2\gamma} \left( \int_0^z dz' g(x,z-z') \p_{z'}^2 \mc{T}_r(z',\tau) + \p_{z'} \mc{T}_r(z',\tau) g(x,z)|_{z' \to 0} - g(x,z-z') \p_{z'}\mc{T}_r(z',\tau)|_{z' \to z}    \right) \nn \\
	=& ~ \frac{1}{2\gamma} \left(  \int_0^z dz' \p_{z'} (g(x,z-z') ) \p_{z'} \mc{T}_r(z',\tau)\right) \nn\\
	=& ~-\frac{1}{\sqrt{\gamma}} \int_0^z dz'  \frac{ e^{-x^2/(4 \gamma (z-z'))} }{\sqrt{4 \pi (z-z') }}\p_{z'} \mc{T}_r(z',\tau) 
}
which gives the relaxation of the correlation fields. The evolution of temperature field is obtained from \eqref{T-C-eq} by putting $x \to 0$ in the above expression for $\mc{C}_r(x,z,\tau)$, we immediately have,
\eqa{\p_\tau \mc{T}_r(z,\tau) = ~\kappa~ \p_z  \int_0^z  \frac{ \p_{z'} \mc{T}_r(z',\tau) }{\sqrt{z-z'}}dz', ~~0\leq z \leq1,~~
	\text{where},~~ \kappa = \frac{1}{\sqrt{ \pi \gamma} }. \label{eqs:T_evo}
}
For infinite system, one can perform the same procedure as done above for finite system using appropriate control parameter and obtain
\eqa{\p_\tau \mc{T}_r(z,\tau) = ~\kappa~ \p_z  \int_{-\infty}^z  \frac{ \p_{z'} \mc{T}_r(z',\tau) }{\sqrt{z-z'}}dz', ~~-\infty\leq z \leq \infty. \label{eqs:T_evo-inf}
}
\subsection{Fractional evolution of temperature in an infinite line}
\noindent
In the previous section we have derived an equation for evolution of temperature the field in an finite system where $y \in(0,1)$.  One can extend the same calculation in an finite system of length $L$ and obtain the same set of bulk equations which now hold for $y \in [0,L]$. We are interested in the behavior of the evolution of temperature profile in  $L \to \infty $ limit, where the effect of boundaries are not important. The evolution equations for the relaxation parts in this case are,
\eqa{
	\gamma \p_{x}^2 \mc{C}_r(x,y,\tau) &= - \p_{y} \mc{C}_r(x,y,\tau) \nn \\
	\p_{y} \mc{T}_r(y,\tau) &= -2 \gamma \p_{x} \mc{C}_r(x,y,\tau)|_{x\to0} \label{eqs:NESSequations}\\
	\p_\tau \mc{T}_r (y,\tau)&=~2\p_y \mc{C}_r(0,y,\tau)\nn
}
where $0\le x\le \infty$ and $0 \le y \le L$. To proceed, we introduce the orthonormal and complete basis in $y \in [0,L]$ , $\phi^\pm_n(y) = \frac{1}{\sqrt{L}}e^{\pm \iu n \pi y/L}$ for $n\ge 1$ and $\phi_0(y) = 1/\sqrt{L}$. Expanding the correlations and temperature in this basis as Fourier series we get,
\eqa{\mc{C}_r(x,y,\tau) = \hat{A}_0(x,\tau) + \sum_{n=1} \hat{A}^+_n(x,\tau) \phi^+_n(y) + \hat{A}^-_n(x,\tau) \phi^-_n(y) \nn \\
	\mc{T}(y,\tau) = \hat{T}_0(\tau) + \sum_{n=1} \hat{T}^+_n(\tau) \phi^+_n(y) + \hat{T}^-_n(\tau) \phi^-_n(y)}
where $\hat{A}_n^{\pm} (x,\tau)= \int_0^L \mc{C}_r(x,y,\tau)  \phi_{n}^\pm (y) dy$, $\hat{A}_0(x,
\tau) = \int_0^L \mc{C}_r(x,y,\tau)  \phi_{0} dy$, $\hat{T}_n^{\pm} (\tau)= \int_0^L \mc{T}_r(y,\tau)  \phi_{n}^\pm (y) dy$, $\hat{T}_0(
\tau) = \int_0^L \mc{T}_r(y,\tau)  \phi_{0} dy$. These  gives the following differential equations for the components,
\eqa{
	\p^2_x\hat{A}^\pm_n(x,\tau) =& ((1 \mp \iu )\alpha_n)^2 \hat{A}^\pm_n(x,\tau), ~ \p^2_x\hat{A}_0(x,\tau) = 0\\
	\pm \frac{\iu n \pi}{L}  \hat{T}^\pm_n(\tau) =&  -2 \gamma \p_x \hat{A}^\pm_n(x,\tau)_{|x\to 0}
	.} where $\alpha_n = \sqrt{\frac{n \pi }{2L\gamma}}$.
The solution to these equations are in general given as,
\eqa{	\hat{A}^\pm_n(x,\tau) =& a^\pm_n(\tau) e^{\pm\alpha_n (1\mp\iu) x} , ~ \hat{A}_0(x,\tau)  =  d(\tau)x +e(\tau) \nn }
We choose solutions which do not blow up at infinity at large $x$ and obey the boundary conditions. We have,
\eqa{
	\mc{C}_r(x,y,\tau) =&  e(\tau) + \sum_{n=1}^\infty a_n^- e^{-\alpha_n (1+\iu) x}\phi_n^-(y) + c.c.
}
where c.c. stands for complex conjugate. $e(\tau)$ is zero because there is no time-dependent source in the system. Using above equations, , we get,
\eqa{ \hat{T}^\pm_n =2\gamma  a_n^\mp(\tau) \alpha_n \frac{(1\mp\iu)}{(n\pi/L)},~
	\dot{\hat{T}}^\mp_n =\mp 2 \iu \frac{n \pi}{L} a_n^\pm(\tau)}
Upon combining these two, we get $	\dot{\hat{T}}_0 = 0$ and 
\eqa{\dot{\hat{T}}^\mp_n &= -\frac{1}{\sqrt{2\gamma}} (1 \pm \iu) \lambda_n^{3/4} \hat{T}^\mp_n,~~n=1,2,3...
}
where  $\lambda_n = (n \pi /L)^2$. This can be interpreted in domain $y \in[0,L]$ as,
\eqa{\p_\tau \mc{T}_r(y,\tau) = -\frac{1}{\sqrt{2\gamma}}(| \Delta|^{3/4} - \nabla  | \Delta|^{1/4} )\mc{T}_r(y,\tau) = -\frac{1}{\sqrt{2\gamma}}\mb{L} \mc{T}_r(y,\tau )}
where $\mb{L}$ is an positive operator defined by its action as, $\mb{L} \phi_n^\pm(y) = \lambda^{3/4}_n (1 - \iu~ \text{sgn}(n) ) \phi_n^\pm(y)$.
With $L \to \infty$ the spectrum becomes continuous as well as the eigenfunctions become plane wave. Thus in infinite system at equilibrium, the evolution of temperature profile is given by a skew-symmetric fractional Laplacian given in Eq.(3) of the main text.

One can alternatively see this equivalence from the integro-differential evolution in infinite space given in 
Eq.~\eqref{eqs:T_evo-inf}. Let us write this equation as 
\eqa{\p_\tau \mc{T}_r(y,\tau) =  \mathcal{L}_{\infty}\mc{T}_r(y,\tau),~\text{where},~\mathcal{L}_{\infty}f(y)=\frac{1}{\sqrt{\pi \gamma}}~ \p_y  \int_{-\infty}^y  \frac{ \p_{y'} f(y') }{\sqrt{y-y'}}dy'.
}
Using the identity
\begin{align}
\int_{-\infty}^ydz \frac{1}{\sqrt{y-z}}~e^{\iu qz}
= \frac{\sqrt{\pi}}{\sqrt{\iu q}}~e^{\iu q y}\nonumber
\end{align}
one can easily show that 
\begin{align}
\mathcal{L}_\infty~e^{\iu  q y} = \lambda_q~e^{\iu q y},
~~\lambda_q =\sqrt{\frac{1}{2\gamma}}[1-\iu~ \text{sgn}(q)]~|q|^{3/2}
\nonumber
\end{align}
which is same as the Fourier spectrum of the skew-symmetric fractional Laplacian given in Eq.(3) of the main text.

\subsection{Series solution of the Fractional PDE in finite domain}
\begin{figure}
	\centering
	\includegraphics[width=0.75\linewidth]{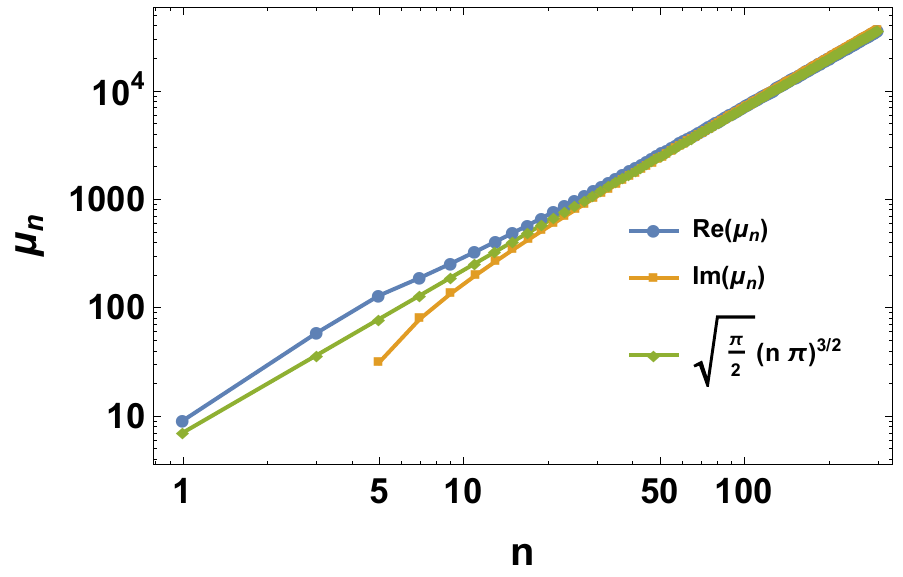}
	\caption{The real  and imaginary part of the alternate eigenvalues for the matrix B. The first $4$ eigenvalues are completely real and distinct. The higher eigenvalue comes in pairs of $\mu_n(1 \pm \iu)$. For large $n$, the eigenvalues are close to $\sqrt{\frac{\pi}{2}} (n \pi)^{3/2}(1 \pm \iu) $. For smaller $n$, there is a deviation from asymptomatic scaling. }
	\label{fig:eigenvalue}
\end{figure}

\begin{figure}
	\centering
	\includegraphics[width=0.85\linewidth]{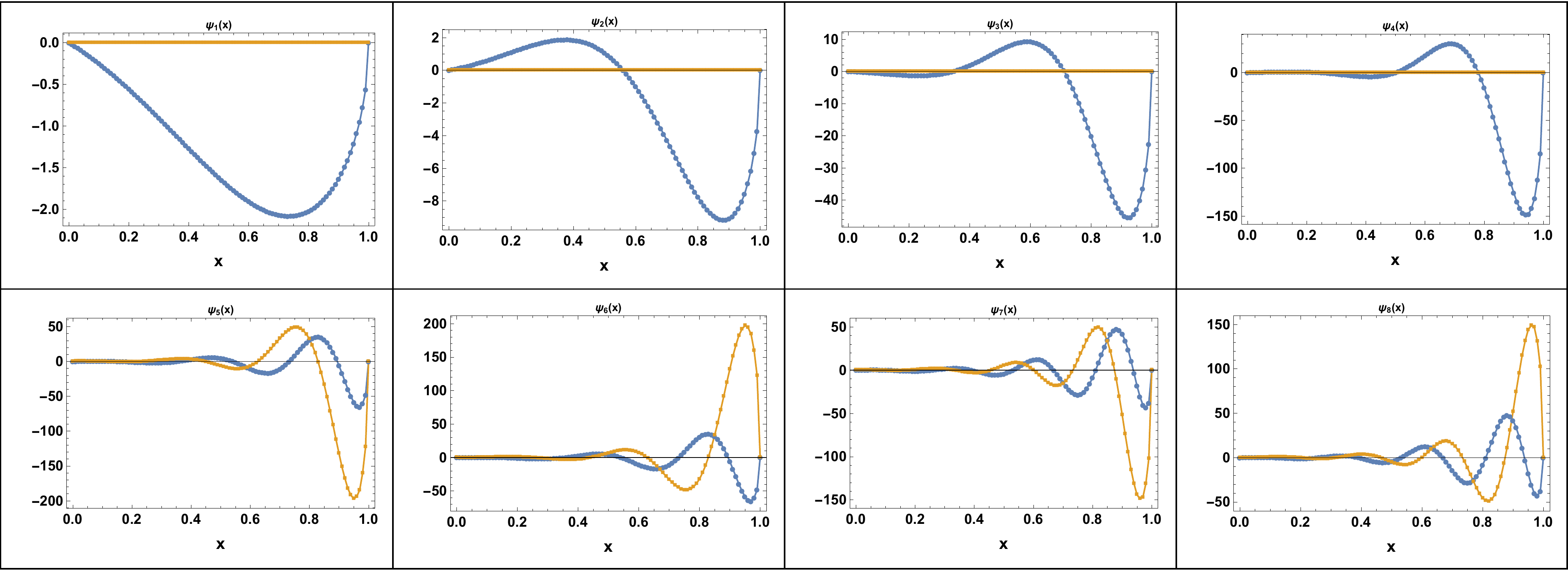}
	\includegraphics[width=0.85\linewidth]{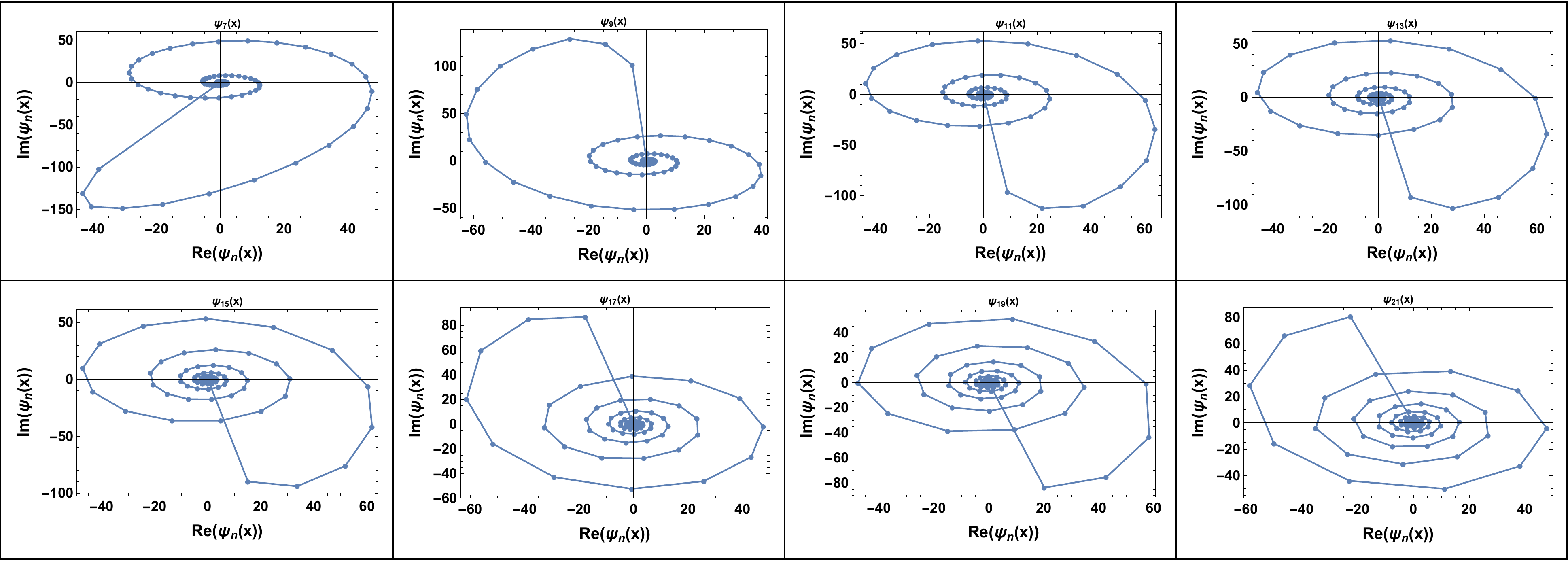}
	\caption{(a)The real (Blue) and Imaginary (Orange) part of the right eigenvectors for the matrix B for the first few eigenvalues.  (b) Polar plots showing the real and imaginary parts of the eigenvectors for $n\ge 7$. The polar plots are for $n =  7 ,9, 11...$. The plots for even ordered eigenvectors  ($N=8,10...$) are related to the eigenvectors of the previous eigenvectors by a reflection around x axis and hence are not plotted. }
	\label{fig:eigenvectors}
\end{figure}

The evolution of the relaxation part of the temperature profile \emph{i.e.} $\mc{T}_r(1-y,\tau) = \mc{T}(y,\tau) - \mc{T}_{ss} (y)$  is given by Eq.~\eqref{eqs:T_evo}. Note that, $\mc{T}_r(z,\tau)$ is zero at both the boundaries: $z=0$ and $z=1$. As a result it natural to expand this function in $\alpha_n(z) = \sqrt{2} \sin (n \pi z),~n =1,2,3...$ complete basis defined in $z \in (0,1)$, as 
$\mc{T}_r(z,\tau) = \sum_n \dx{\theta}_n (\tau)\alpha_n(z)$. Substituting this form in  Eq.~\eqref{eqs:T_evo}, we have,
\eqa{ \sum_n \dot{\dx{\theta}}_n \alpha_n(z) = \kappa~ \sum_n \dx{\theta}_n (\tau) (n \pi) \p_z  \int_0^z \frac{ \phi_n(z')}{\sqrt{ z-z'}} dz'}
Now let us expand the function $f_n(z) = \p_z \int_0^z  \frac{ \phi_n(z')}{\sqrt{ z-z'}} dz' $ also in orthogonal basis $\alpha_n(y),~n=1,2....$. Let the expansion is given as $f_n(z) = \sum_{l=1} \zeta_{nl} \alpha_l(z)$ where 
$\zeta_{nl}=\int_0^1dz~f_n(z)~ \alpha_l(z)$. As a result we have, 
\eqa{ \sum_{n=1} \dot{\dx{\theta}}_n \alpha_n(z) = \kappa~ \sum_{n,l=1} \dx{\theta}_n (\tau)(n \pi) \zeta_{nl} \alpha_l(z)}
Using orthogonality, this can be written in vector notation as ($\dx{\theta}_n = \iprod{n}{\dx{\theta}}$),
\eqa{\ket{\dot{\dx{\theta}}} = \kappa~ \mathbf{B}\ket{\dx{\theta}} ,} where $B_{nk} = (n \pi)\zeta_{nk} $.
If $R$ is the matrix which diagonalizes $\mathbf{B}$ as $R^{-1}\mathbf{B}R = \Lambda$, then the time dependent solution is given as $\ket{\dx{\theta}(\tau)} = R e^{\kappa \Lambda  \tau}R^{-1}  \ket{\dx{\theta}(0)}$ and temperature at time $\tau$ is given as $\mc{T}(y,\tau) = \mc{T}_{ss}(y)+ \iprod{\alpha(1-y)}{\dx{\theta}(\tau)}$. The eigenvalues of the bounded skew-fractional laplacian, $\mathbf{B}$ have interesting behavior, the first four of them are real and distinct. The higher eigenvalues all come in complex conjugate pairs of form $\mu_n ( 1 \pm \iu)$. For large $n$, $\mu_n$ goes as $\sqrt{\frac{\pi}{2}}(n \pi)^{3/2}$, but for smaller $n$ there is a systematic deviation due to the effect of finite domain. In Fig. \ref{fig:eigenvalue}, the real part of alternate eigenvalues are plotted as a function of $n$, where the asymptotic scaling with $\sqrt{\frac{\pi}{2}}(n \pi)^{3/2}$ is seen clearly for large $n$.  The eigenvectors of the operator is defined as  $\psi_n(y) = \sum_{l=1} R^{-1}_{nl} \alpha_l(y)$. Numerically computing this gives, the first six eigenvectors to be completely real. The eigenvectors corresponding to higher eigenvalues are complex and and comes in pairs. The real and imaginary parts of the first few eigenvectors are shown in Fig.\ref{fig:eigenvectors}(a). In Fig.\ref{fig:eigenvectors}(b), the real and the imaginary part of the eigenvector for $n\ge 7$ is plotted in polar plots. For plane wave solutions these would have been circles of length 1,  here the polar plot shows a spiral decay to origin owing to the skewness of the operator.
As the temperature field evolves at much faster timescales compared to the correlation field, the time dependent solution for correlations $\mc{C}_r(x,1-y,\tau)  $ is governed by the evolution of the temperature field. The  solution for evolution of correlations is written as, $ \mc{C}(x,y,\tau)  = \mc{C}_r(x,1-y,\tau) + \mc{C}_{ss} (x,y)$ , where,
\eqa{\mc{C}(x,z,\tau) =& - \int_0^z dz' \frac{e^{-x^2/(4 \gamma (z-z'))}}{\sqrt{4 \pi \gamma (z-z')}} \p_{z'} T_r(z',\tau) \nn \\
	=& \sum_{n=1} \dx{\theta}_n(\tau) (n \pi) \int_0^z  dz' \frac{e^{-x^2/(4 \gamma (z-z'))}}{\sqrt{4 \pi \gamma (z-z')}} \phi_n(z')
}
where, $\phi_n(y) = \sqrt{2} \cos(n \pi y),~n\ge 1$ and $\phi_0(y) = 1$. The  integral can be evaluated explicitely and doing the summations gives the evolution of the correlation fields. 
\end{document}